\documentclass[prl,aps,tightenlines,preprint,superscriptaddress]{revtex4-1}

\usepackage{graphicx}
\usepackage{amsmath}
\usepackage{color}
\usepackage{ulem}

\newcommand{\cev}[1]{\reflectbox{\ensuremath{\vec{\reflectbox{\ensuremath{#1}}}}}}

\begin{document}
	
\title{Nucleon decay in the deuteron}
\author{F. Oosterhof}
\affiliation{Van Swinderen Institute for Particle Physics and Gravity,
		University of Groningen, 9747 AG Groningen, The Netherlands}
	
\author{J. de Vries}
\affiliation{Institute for Theoretical Physics Amsterdam and Delta Institute for Theoretical Physics, University of Amsterdam, Science Park 904, 1098 XH Amsterdam, The Netherlands}
\affiliation{Nikhef, Theory Group, Science Park 105, 1098 XG, Amsterdam, The Netherlands}
\affiliation{Amherst Center for Fundamental Interactions,
		Department of Physics, University of Massachusetts Amherst,
		Amherst, Massachusetts 01003, USA}
\affiliation{RIKEN BNL Research Center, Brookhaven National Laboratory,
		Upton, New York 11973-5000, USA}
	
\author{R. G. E. Timmermans}
\affiliation{Van Swinderen Institute for Particle Physics and Gravity,
		University of Groningen, 9747 AG Groningen, The Netherlands}
	
\author{U. van Kolck}
\affiliation{Universit\'e Paris-Saclay, CNRS/IN2P3, IJCLab, 91405 Orsay, France}
\affiliation{Department of Physics, University of Arizona,
		Tucson, Arizona 85721, USA}
	\vspace{1cm}	
\begin{abstract}		
We calculate the lifetime of the deuteron from dimension-six quark operators 
that violate baryon number by one unit. 
We construct an effective field theory (EFT) for $|\Delta B|=1$ interactions 
that give rise to nucleon and $\Delta B=1$ deuteron decay in a systematic 
expansion. 
Nucleon decay introduces imaginary parts in the low-energy constants of the 
$\Delta B=0$ nuclear interactions in Chiral EFT.
We show that up to and including next-to-leading order the deuteron decay rate 
is given by the sum of the decay rates of the free proton and neutron.
The first nuclear correction is expected to contribute at the few-percent level
and comes with an undetermined low-energy constant. 
We discuss its relation to earlier potential-model calculations.	
\end{abstract}
	
\date{\today}
	
	
\maketitle	

In the Standard Model (SM) of particle physics the observed stability of 
the proton is attributed to an accidental global $U(1)$ symmetry. 
The associated, classically conserved quantity, baryon number $B$,
is, however, 
broken by small quantum effects that preserve $B-L$,
where $L$ denotes lepton number. 
If it is not associated 
to a local, gauge symmetry, 
$B$ is expected to be violated in extensions of the SM to higher energies. 
In fact, proton decay is an outstanding prediction of Grand Unified Theories, 
in which the strong, weak, and electromagnetic interactions unify at an 
ultrahigh scale of the order of $10^{16}$ GeV \cite{Babu:2013jba}. 
At present, the lower bounds on the proton lifetime for the two important decay
channels $p\rightarrow\pi^0+e^+$ and $p\rightarrow\pi^++\bar{\nu}$ are 
$1.6 \cdot 10^{34}$ y and $3.9\cdot 10^{32}$ y at 90\% confidence level, 
respectively \cite{Miura:2016krn,Abe:2013lua}. 
Such processes violate baryon number by one unit ($\Delta B=1$).
 
Experiments that search for nucleon decay aim to detect it inside nuclei. 
The lifetime of a nucleon bound in a nucleus may differ from that of a 
free nucleon due to nuclear 
interactions, which make theoretical estimates difficult. 
We examine here the effect of these
interactions on the lifetime of the deuteron, the simplest nucleus consisting 
of more than one nucleon, in effective field theory (EFT). 
EFT allows for systematic and model-independent calculations of 
low-energy processes. Recently, EFT was applied to processes in which 
baryon number is violated by two 
units \cite{FemkeMSc,Bijnens:2017xrz,Oosterhof:2019dlo,Haidenbauer:2019fyd}, 
in particular neutron-antineutron oscillations in free space and in the
deuteron. We follow Ref.~\cite{Oosterhof:2019dlo}, in which the 
$\Delta B=2$ deuteron decay rate was calculated in a systematic expansion.

\begin{figure}[tb]
\centering
\includegraphics[scale=1.0]{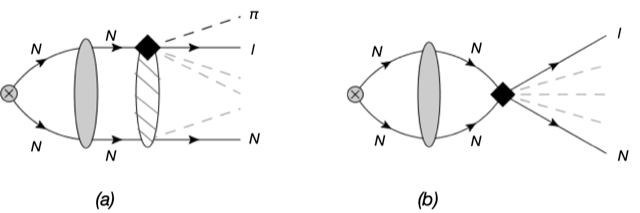}
\caption{The two classes of deuteron decay. 
The crossed circle denotes the deuteron, 
the solid line a nucleon ($N$) or a lepton ($l$), 
the dashed black line a pion ($\pi$), 
and the gray dashed lines denote possible more pions. 
The black diamond stands for a $\Delta B=1$ interaction, 
while the gray blob represents the propagation and strong interaction
of two nucleons,
and the dashed blob the possibility of additional 
pion interactions.  
\label{Fig1}}
\end{figure}

We restrict ourselves to the lowest-dimension $|\Delta B|=1$ operators 
in the SM EFT that satisfy the full SM gauge symmetry and construct an EFT 
with pions and nucleons describing nucleon and deuteron decay.
These operators mediate the decay of the 
deuteron to a nucleon, an antilepton, and one or more mesons. We distinguish 
between two classes of 
$\Delta B=1$ deuteron decay, which are illustrated in Fig.~\ref{Fig1}. 
In processes of the first class, Fig.~\ref{Fig1}(a), one of the nucleons 
decays to an antilepton and one or more mesons, while it may interact with 
the other nucleon via long-range interactions. In this class, we can directly 
relate the deuteron decay rate to 
$\Delta B=1$ decay in the one-nucleon sector via known SM physics. 
In processes of the second class, Fig.~\ref{Fig1}(b),
the two nucleons in the deuteron are converted to an antilepton and 
a nucleon and possibly mesons via a short-range $\Delta B=1$ interaction. 
Since the produced mesons and antileptons have high momenta, they cannot be 
treated explicitly in the EFT. We therefore introduce imaginary parts in the 
low-energy constants of $\Delta B=0$ interactions, which generate 
inclusive decay widths.
We show that the power counting 
--- that is, the ordering of the EFT
interactions according to the expected magnitude of their contributions
to observables ---
dictates that deuteron decay is dominated by free-nucleon decay,
which is described by the first class.
The first nuclear correction is expected to come from 
the second class at the few-percent level.

The lowest-dimension $|\Delta B|=1$ operators that satisfy the full 
$SU(3)_c\otimes SU(2)_L\otimes U(1)_Y$ SM gauge symmetry are 
dimension-six operators constructed from three quark fields and 
one lepton field. We  focus on the operators with $u$- and $d$-quark fields
and at most one $s$-quark field, because conservation of energy does not 
allow the deuteron to decay via a $\Delta B = 1$ process 
to a final state with two strange quarks or two strange antiquarks. 
The Lagrangian of the $|\Delta B|=1$ dimension-six operators 
can be written as  
\cite{Weinberg:1979sa,Wilczek:1979hc,Abbott:1980zj,Claudson:1981gh}
\begin{eqnarray}
{\cal L}_{|\Delta B|=1}= \sum_{i=1}^4\sum_{d=1}^2 {\cal C}_d^{(i)} {\cal Q}_d^{(i)} 
+ \sum_{i=1}^6\sum_{d=1}^2 \tilde{\cal C}_d^{(i)} \tilde{\cal Q}_d^{(i)}   
+ \text{H.c.} \ ,
\end{eqnarray}
where ${\cal C}_d^{(i)}$ and $\tilde{\cal C}_d^{(i)}$ are Wilson coefficients 
that depend on physics beyond the SM, 
${\cal Q}_d^{(i)}$ are operators with $u$- and $d$-quark fields,
\begin{eqnarray}
\label{eqOpsud}
{\cal Q}^{(1)}_d&=& ({d_R}^T_\alpha C  {u_R}_\beta)
\left[({u_L}^T_\gamma C {e_L}_d) - ({d_L}^T_\gamma C {\nu_L}_d)\right]
\varepsilon_{\alpha\beta\gamma} \ ,
\nonumber\\
{\cal Q}^{(2)}_d&=& ({d_L}^T_\alpha C  {u_L}_\beta)({u_R}^T_\gamma C {e_R}_d)
\varepsilon_{\alpha\beta\gamma} \ ,
\nonumber\\
{\cal Q}^{(3)}_d&=& ({d_L}^T_\alpha C  {u_L}_\beta)
\left[({u_L}^T_\gamma C {e_L}_d) - ({d_L}^T_\gamma C {\nu_L}_d)\right]
\varepsilon_{\alpha\beta\gamma} \ ,\nonumber\\
{\cal Q}^{(4)}_d&=& ({d_R}^T_\alpha C  {u_R}_\beta)({u_R}^T_\gamma C {e_R}_d)
\varepsilon_{\alpha\beta\gamma} \ ,
\end{eqnarray}
and $\tilde {\cal Q}_d^{(i)}$ operators with one strange-quark field,
\begin{eqnarray}
\label{eqOpsuds}
\tilde {\cal Q}^{(1)}_d&=& ({s_R}^T_\alpha C  {u_R}_\beta)
\left[({u_L}^T_\gamma C {e_L}_d) - ({d_L}^T_\gamma C {\nu_L}_d)\right]
\varepsilon_{\alpha\beta\gamma} \ ,
\nonumber\\
\tilde {\cal Q}^{(2)}_d&=& ({s_L}^T_\alpha C  {u_L}_\beta)({u_R}^T_\gamma C {e_R}_d)
\varepsilon_{\alpha\beta\gamma} \ ,
\nonumber\\
\tilde {\cal Q}^{(3)}_d&=& ({s_L}^T_\alpha C  {u_L}_\beta)
\left[({u_L}^T_\gamma C {e_L}_d) - ({d_L}^T_\gamma C {\nu_L}_d)\right]
\varepsilon_{\alpha\beta\gamma} \ ,
\nonumber\\
\tilde {\cal Q}^{(4)}_d&=& ({s_R}^T_\alpha C  {u_R}_\beta)({u_R}^T_\gamma C {e_R}_d)
\varepsilon_{\alpha\beta\gamma} \ ,
\nonumber\\
\tilde {\cal Q}^{(5)}_d&=& ({d_R}^T_\alpha C  {u_R}_\beta)({s_L}^T_\gamma C {\nu_L}_d)
\varepsilon_{\alpha\beta\gamma}\ ,
\nonumber\\
\tilde {\cal Q}^{(6)}_d&=& ({d_L}^T_\alpha C  {u_L}_\beta)({s_L}^T_\gamma C {\nu_L}_d)
\varepsilon_{\alpha\beta\gamma} \,.
\end{eqnarray}
In these operators 
the subscript $d=1,2$ indicates the lepton generation, 
the subscript $L,R$ indicates the chirality of the fermion field, 
$C$ is the charge-conjugation matrix,  
$\alpha,\beta,\gamma$ are color indices, 
and $\varepsilon_{\alpha\beta\gamma}$ is the Levi-Civita tensor.
The Wilson coefficients are expected to be suppressed by the scale of 
new physics $\Lambda_{|\Delta B|=1}$, \textit{i.e.} 
${\cal C}_d^{(i)}={\cal O}(c_d^{(i)} \Lambda_{|\Delta B|=1}^{-2})$ and 
$\tilde{\cal C}_d^{(i)}={\cal O}(\tilde c_d^{(i)} \Lambda_{|\Delta B|=1}^{-2})$, 
where $c_d^{(i)}$ and $\tilde c_d^{(i)}$  are dimensionless constants. 
We use ${\cal C}$ to denote 
the values of ${\cal C}_d^{(i)}$ and $\tilde{\cal C}_d^{(i)}$.

At low energies QCD is non-perturbative, which complicates the calculation of 
the effects of these interactions on observables.
We therefore use Chiral EFT ($\chi$EFT) \cite{Weinberg:1978kz,Weinberg:1990rz},
which is a low-energy EFT of QCD with nucleons and pions as the relevant 
degrees of freedom. $\chi$EFT is based on the approximate 
$SU(2)_L\otimes SU(2)_R$ chiral symmetry of QCD, spontaneously broken to 
the isospin subgroup $SU(2)_I$ and 
explicitly broken by the $u$- and $d$-quark masses.
The pions appear naturally in the theory as 
pseudo-Goldstone bosons with a mass well below the 
chiral-symmetry-breaking scale,
$m_\pi\simeq 140$ MeV $\ll \Lambda_\chi \sim 2\pi F_\pi$,
where $F_\pi \simeq 185$ MeV the pion decay constant.
The chiral Lagrangian is constructed from all terms consistent with 
the symmetries of QCD and ordered in the number of derivatives and 
quark masses. Observables at typical momenta $Q \lesssim m_\pi$
are calculated in an expansion in $Q/\Lambda_\chi$.
Since nucleons have a mass $m_N\simeq 940$ MeV $\sim \Lambda_\chi$,
they are treated as non-relativistic in the regime of validity of the theory. 
Terms in the chiral Lagrangian contain the so-called low-energy constants
(LECs), which have to be calculated with non-perturbative methods, 
in particular lattice QCD, 
or determined from experiment.
Under an assumption of naturalness, they can be estimated by
naive dimensional analysis (NDA) \cite{Manohar:1983md}.
(For a review of the form and limitations of NDA in nuclear systems,
see Ref. \cite{vanKolck:2020plz}.)

We first discuss the one-nucleon sector. The Lagrangian for non-relativistic 
nucleon fields $N = (p\,n)^T$ interacting with pion fields  $\pi^a$ ($a=1,2,3$)
is 
\begin{eqnarray}\label{eqL2B0}
{\cal L}_N &=&
N^\dagger \left(i \partial_0+\frac{\nabla^2}{2m_N}\right) N 
-\frac{1}{2}\,\pi^a\left(\partial^2+m_\pi^2\right)\pi^a 
+\frac{g_A}{F_\pi}\,{N^\dagger}\sigma_k\tau^aN\,\nabla_k\pi^a
\nonumber\\
&&+\frac{1}{2}N^\dagger\left(\alpha_0+ \alpha_1 \tau^3 \right) N 
+ \frac{1}{F_\pi}N^\dagger \left[\alpha_2\left(\tau^a - \delta^{a3}\tau^3\right)
+\alpha_3\,\varepsilon^{3ab}\tau^b \right] N \,\pi^a
+ \dots\ ,
\end{eqnarray}
where    
$\sigma_k$ ($k=1,2,3$) are Pauli spin matrices, 
$\tau^a$ ($a=1,2,3$) are Pauli isospin matrices,
and $\varepsilon^{abc}$ is the Levi-Civita tensor in isospin space.
Here and below the dots denote terms that do not contribute to the order 
of our calculation. 
The first line in Eq. (\ref{eqL2B0}) is the standard $B$-conserving Lagrangian,
with $g_A\simeq 1.27$ the axial-vector coupling constant.
Before they are expanded in powers of $\pi^2/F_\pi^2$, 
these operators transform as tensor products of scalars and vectors under 
$SO(4)\sim SU(2)_L\otimes SU(2)_R$, reflecting the
pattern of explicit chiral-symmetry breaking in the QCD Lagrangian. 
The terms with the complex LECs $\alpha_{0,1,2,3}$ 
are induced by $|\Delta B|=1$ physics and 
contribute to the inclusive proton and neutron decay rate. 
These operators, before they are expanded in powers of pion fields, 
transform under chiral symmetry 
as the Kronecker products of each combination of two $|\Delta B|=1$ operators 
in Eqs. \eqref{eqOpsud} and \eqref{eqOpsuds} that have zero net baryon number, 
lepton number, and strangeness. 
From NDA we expect
\begin{equation}\label{eqNDAalpha}
\alpha_{0,1,2,3}
= {\cal O}\left(\frac{\Lambda_\chi^5 {\cal C}^2}{(4\pi)^4}\right) 
\ . 
\end{equation}
We do not write explicit $|\Delta B|=1$ interactions, 
because the mesonic final states 
in Fig.~\ref{Fig1} contain hard pions with momenta outside the regime of 
validity for $\chi$EFT. 
Instead, we calculate the imaginary part of the pole in the nucleon and 
deuteron propagators to determine their inclusive widths. 
These imaginary parts stem from the
imaginary parts of the newly introduced LECs, which therefore enable
nucleon and nuclear decay. 

The fully dressed propagator for a nucleon $i=p,n$ of energy $E$ and
three-momentum $\vec{p}$ can be written in the form
\begin{equation}
G_i(E, \vec{p}\,) =\frac{i Z_{i}}{E -\vec{p}\,^{2}/2m_N +\ldots + i\Gamma_i/2 }
\ ,
\end{equation}
where 
$Z_{i}$ is the wavefunction renormalization factor, and 
\begin{eqnarray}\label{eqGamp}
\Gamma_p &=& \mathrm{Im} (\alpha_0 + \alpha_1) +\dots\ ,
\\
\label{eqGamn}
\Gamma_n &=& \mathrm{Im}(\alpha_0 - \alpha_1) + \dots
\end{eqnarray}
are the proton and neutron decay rates, respectively. 
The dots indicate higher-order corrections coming from loop diagrams and 
contact interactions with 
insertions of the quark masses and/or derivatives.
These corrections start at relative ${\cal O}(Q^2/\Lambda_\chi^2)$ and 
are discussed in detail in a separate paper \cite{toappear}.  
From the NDA estimate in Eq. (\ref{eqNDAalpha}) we have
\begin{eqnarray}\label{eqGammaest}
\Gamma_{p,n}={\cal O}\left(\frac{\Lambda_\chi^5 {\cal C}^2}{(4\pi)^4}\right)
\sim 10^{-4} \; {\cal C}^2 \,\text{GeV}^5  \ .
\end{eqnarray}
This estimate of the inclusive proton and neutron decay rate is consistent with
lattice QCD calculations of the decay 
to a meson and a positron or antimuon, where the largest contribution to the 
decay rate is 
$\simeq 3\cdot 10^{-4}\;{\cal C}^2\,\text{GeV}^5$  
\cite{Aoki:2017puj}. 

To calculate the deuteron decay rate we need to include two-nucleon ($N\!N$) 
interactions. The scattering length in the ${}^3S_1$ channel is unnaturally 
large, which is related to the small binding momentum of the deuteron,
$\kappa = \sqrt{m_N B_d}\simeq 45$ MeV, with $B_d\simeq2.225$ MeV the deuteron 
binding energy. 
This translates in $\chi$EFT to $N\!N$  LECs with magnitudes larger than 
expected by NDA \cite{Kaplan:1998tg,Bedaque:2002mn}. 
With this enhanced scaling the leading-order (LO) $N\!N$ interaction 
must be iterated to all orders, resulting in an $S$-matrix pole
associated with the deuteron bound state. 
Subleading 
contact interactions and pion exchange between nucleons can be treated 
perturbatively in an expansion $Q/\Lambda_{N\!N}$, where 
$Q\sim m_\pi\sim \kappa$ and $\Lambda_{N\!N}\equiv 
4\pi F_\pi^2/g_A^2 m_N\sim F_\pi$. 
The same scheme has been used to successfully compute the electromagnetic 
form factors of the deuteron 
\cite{Kaplan:1998sz,Savage:1999cm,deVries:2011re,Mereghetti:2013bta}
and $N\!N$ scattering up to center-of-mass momenta around 100 MeV
\cite{Fleming:1999ee}.

The Lagrangian for $N\!N$ contact interactions is
\begin{eqnarray}\label{eqL4B0}
{\cal L}_{N\!N} &=& -\left(C_0+D_2m_\pi^2\right)
\left(N^TP_i N\right)^\dagger\left(N^TP_i N\right)
\nonumber \\
&&+ \frac{C_2}{8}\left\{\left(N^TP_i N\right)^\dagger
\left[N^TP_i (\vec{\nabla} - \cev{\nabla})^2 N\right] +\text{H.c.}\right\}
+ \dots \ ,
\end{eqnarray}
where $P_i \equiv\sigma_2\sigma_i\tau^2/\sqrt{8}$
projects an $N\!N$ pair onto the isospin-singlet
${}^3S_1$ state and $C_0$, $D_2$, and $C_2$ are LECs. 
The operator with the LEC $\mathrm{Re}\,C_0 = {\cal O}(4\pi/m_N\kappa)$
is the LO interaction that is iterated to all orders.
It produces a bound state with binding momentum $\kappa$ 
if \cite{Kaplan:1998tg}
\begin{eqnarray}
\label{eqReC0}
\mathrm{Re}\,C_0 = \frac{4\pi}{m_N(\kappa-\mu)} +\dots \ ,
\end{eqnarray}
where $\mu$ is the renormalization scale. 
$\mathrm{Re}\,C_2\sim \mathrm{Re}\,D_2 = {\cal O}(4\pi/m_N\kappa^2\Lambda_{N\!N})$
are LECs contributing to $N\!N$ scattering at next-to-leading order (NLO)
in the $Q/\Lambda_{N\!N}$ expansion. 
The imaginary part of the $N\!N$ 
LECs is produced by two insertions of $|\Delta B|=1$ physics. 
With a $\Lambda_{N\!N}^2/\kappa^2$ enhancement over NDA 
due to renormalization by the LO $N\!N$ interaction,
\begin{eqnarray}
\label{eqImC0}
\mathrm{Im}\, C_0 = {\cal O}\left(\frac{\Lambda_{N\!N}^4{\cal C}^2}{\kappa^2}
\right) \,.
\end{eqnarray}
By requiring $N\!N$ scattering amplitudes to be independent of the 
renormalization scale, we find that $\mathrm{ Im}\,C_0$ 
must satisfy the renormalization-group equation
\begin{eqnarray}
\frac{\rm d}{{\rm d}\mu}\mathrm{Im}\, C_0 = \frac{m_N}{2\pi}\mathrm{Re}(C_0)\,
\mathrm{Im}\, C_0
+ \dots \, .
\label{ImC0run}
\end{eqnarray}

The deuteron decay rate is obtained from the imaginary part of the deuteron 
propagator. Following Ref.~\cite{Kaplan:1998sz}, the propagator of a deuteron 
with four-momentum 
$p^\mu=(2m_N+\vec p\,^2/4m_N +\bar E +\ldots,\vec{p}\,)$ can be written in terms
of the irreducible two-point function $\Sigma(\bar E)$, which consists of all 
diagrams that do not fall apart when cutting any $\mathrm{Re}\, C_0$ vertex, 
\begin{eqnarray}
G_d(\bar{E}) = \frac{\Sigma(\bar{E})}{1+i\, \mathrm{Re}(C_0) \,\Sigma(\bar{E})}
= \frac{iZ_d}{\bar{E} + B_d + i\Gamma_d/2} + \dots \, ,
\label{deuteronProp}
\end{eqnarray}
where we expanded around $\bar{E} = -B_d$.
$Z_d$ is the deuteron wavefunction renormalization factor and 
\begin{eqnarray}
\Gamma_d = \left.\frac{2\, \mathrm{Im}(i\Sigma(\bar E))}
{\mathrm{Re}({\rm d}i\Sigma(\bar E)/{\rm d}\bar E)}\right|_{\bar E = -B_d}
+ \dots
\label{optical}
\end{eqnarray}
is the deuteron decay rate. 
Up to NLO \cite{Kaplan:1998sz},
\begin{eqnarray}
\mathrm{Re}\left(\frac{{\rm d}i\Sigma(\bar E)}{{\rm d}\bar E}\right)
\bigg{|}_{\bar E = -B_d} &=&
\frac{m_N^2}{8\pi\kappa}
\bigg\{1 + \frac{m_N}{2\pi}(\kappa-\mu)
\left[ \kappa(\mu - 2\kappa) \, \mathrm{Re}\,C_2 
+ m_\pi^2\, \mathrm{ Re}\,D_2 \right]
\nonumber\\
&&+\frac{2}{\Lambda_{N\!N}}\left(\kappa - \mu + 
\frac{m_\pi^2}{m_\pi+2\kappa}
\right)
\bigg\}\ .
\label{denom}
\end{eqnarray}

\begin{figure}[t]
\centering
\includegraphics[scale=0.55]{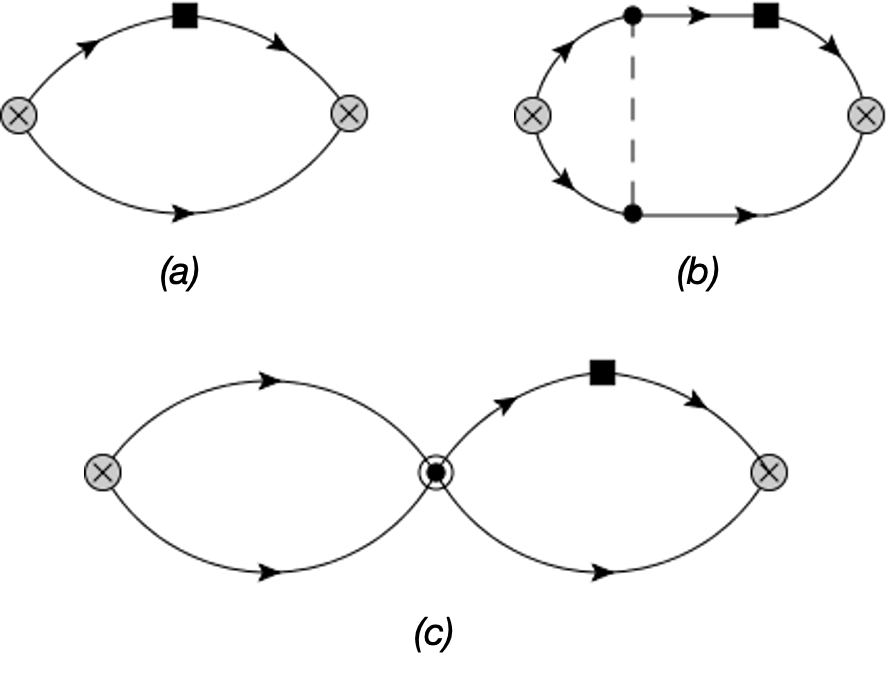}
\caption{Diagrams contributing to the deuteron decay rate up to NLO. 
The solid square denotes an $\mathrm{Im}\,\alpha_0$ vertex, 
the solid circle a $g_A$ vertex, and 
the encircled circle a $\mathrm{Re}\,C_2$ or $\mathrm{Re}\,D_2$ vertex.
Other notation as in Fig. \ref{Fig1}.
\label{Fig2}}
\end{figure}

The diagrams that contribute to $\mathrm{ Im}(i\Sigma(\bar{E}))$ up to NLO 
are shown in Fig. \ref{Fig2}. 
Each loop counts as $Q^5/4\pi m_N$, each nucleon propagator as 
$m_N/Q^2$, and each pion propagator as $1/Q^2$.
The diagrams in Fig. \ref{Fig2} describe deuteron decay as 
illustrated in Fig. \ref{Fig1}(a).
Figure \ref{Fig2}(a) contributes at LO  
and Fig. \ref{Fig2}(b,c) at NLO. 
They amount to
\begin{eqnarray}
\mathrm{Im}(i\Sigma(-B_d)) 
&=& \frac{m_N^2}{8\pi \kappa} \, \mathrm{Im}\,\alpha_0\,
\bigg\{1 + \frac{m_N}{2\pi} (\kappa-\mu) 
\left[\kappa(\mu - 2\kappa) \, \mathrm{Re}\,C_2 
+ m_\pi^2 \, \mathrm{Re}\,D_2\right] 
\nonumber\\
&&+ \frac{2}{\Lambda_{N\!N}}\left(\kappa-\mu+\frac{m_\pi^2}{m_\pi + 2\kappa} 
\right)
\bigg\} \ .
\label{num}
\end{eqnarray}
Substituting Eqs. (\ref{denom}) and (\ref{num}) into Eq. (\ref{optical}) 
we find for the decay rate up to NLO
\begin{equation}\label{eqGammd}
\Gamma_d = 2\,\mathrm{Im}\,\alpha_0
= {\Gamma_p + \Gamma_n}  \ , 
\end{equation}
where in the last equality we used Eqs. \eqref{eqGamp} and \eqref{eqGamn}.

The contributions to $\Gamma_d$ from the NLO diagrams in Fig. \ref{Fig2}(b,c)  
cancel against the $\mathrm{Re}\,C_2$, $\mathrm{Re}\,D_2$, and pion terms 
in Eq. \eqref{denom}. 
This cancellation is not accidental. 
The derivative with respect to $\bar E$ acts on the nucleon propagators and 
effectively adds an additional propagator with the same momentum. 
The resulting term then cancels against the corresponding diagrams contributing
to Fig. \ref{Fig2}. This cancellation also works at higher orders. 
For example,
additional insertions in Fig. \ref{Fig2} 
of $\mathrm{Re}\,C_2$ and $\mathrm{Re}\,D_2$ vertices and/or pion exchanges 
proportional to $g_A^2$ 
before or after the $\mathrm{Im}\,\alpha_{0}$ vertex 
cancel against similar corrections to the two-point function 
in $\mathrm{Re}({\rm d}i\Sigma(\bar E)/{\rm d}\bar E)|_{\bar E = -B_d}$.

\begin{figure}[t]
\centering
\includegraphics[scale=1.0]{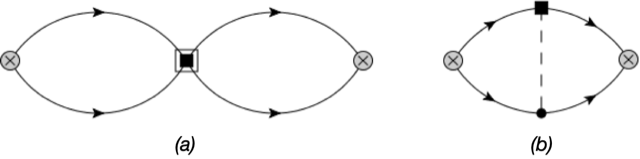}
\caption{Higher-order diagrams contributing to the deuteron decay rate. 
The squared square denotes an $\mathrm{Im}\,C_0$ vertex.
Other notation as in Figs. \ref{Fig1} and \ref{Fig2}.
\label{Fig3}}
\end{figure}

The first non-zero correction to the deuteron decay rate 
arises from the diagram shown in Fig.~\ref{Fig3}(a). 
This diagram describes deuteron decay via a short-range interaction between 
the nucleons, as illustrated in Fig. \ref{Fig1}(b). 
Following the power counting rules and using the NDA estimate for 
$\mathrm{Im}\,C_0$ in Eq. (\ref{eqImC0}), it is expected 
to contribute at relative 
${\cal O}(\kappa \Lambda_{N\!N}/\Lambda_\chi^2) \sim 10^{-2}$.

Long-range nuclear effects in which one or more low-energy pions are emitted 
from the vertex describing nucleon decay can also affect the deuteron decay 
rate. The lowest-order diagram of this type is shown in Fig. \ref{Fig3}(b),
which is expected 
at relative ${\cal O}(\kappa^2/\Lambda_\chi\Lambda_{N\!N}) \sim 10^{-2}$
--- that is, comparable to the diagram in Fig. \ref{Fig3}(a). 
However, the diagram in Fig. 3(b) vanishes: 
the integrand of the loop integral is odd in three-momentum 
because the nucleon decay vertex emitting a pion has no derivative 
(see Eq. (\ref{eqL2B0})) while the $g_A$ vertex has one spatial derivative. 
Replacing one of the pion-nucleon vertices by a subleading vertex with one
additional derivative leads to a suppression of ${\cal O}(\kappa/\Lambda_\chi)$.
Long-range nuclear effects affecting the deuteron decay rate 
are therefore expected
at ${\cal O}(\kappa^3/\Lambda_\chi^2\Lambda_{N\!N})\sim 10^{-3}$ or higher. 

The part of the Lagrangian in Eq. (\ref{eqL2B0}) that comes from 
$|\Delta B| = 1$ physics has at the next order terms with a time derivative 
that cannot be removed by a field redefinition without the appearance
of time-derivative terms beyond LO in the $B$-conserving part \cite{toappear}. 
The corrections to the deuteron decay rate due to these terms appear at 
relative ${\cal O}(\kappa^2/\Lambda_\chi^2)$ and are thus
of higher order than our estimate for the correction from the diagram 
in Fig. \ref{Fig3}(a).

We conclude that the deuteron decay rate is dominated by the sum of the proton 
and neutron decay rates, $\Gamma_p$ and $\Gamma_n$, with a correction at
relative ${\cal O}(\kappa \Lambda_{N\!N}/\Lambda_\chi^2)$, {\it viz.}
\begin{equation}\label{eqGammdmore}
\Gamma_d 
= {\Gamma_p + \Gamma_n}  
- \frac{\kappa}{\pi}(\kappa-\mu)^2 \, {\mathrm{Im}\,C_0} \ ,
\end{equation}
where further corrections are expected at relative
${\cal O}(\kappa^2/\Lambda_\chi^2)$.
The contribution from the LEC $\mathrm{Im}\,C_0$ is independent of the 
renormalization scale $\mu$, as can be seen from Eq. \eqref{ImC0run}.
It represents the first nuclear correction to the deuteron decay rate.
It is presently unknown, but it is expected to be small, of the order of a 
few percent. 

That nucleon decay rates dominate the deuteron decay rate agrees with 
earlier, model-based calculations \cite{Dover:1981zj,AlvarezEstrada:1982wg}. 
However, Refs. \cite{Dover:1981zj,AlvarezEstrada:1982wg} found larger nuclear 
corrections than we do.
The reason for this discrepancy is not clear because the theoretical frameworks
are very different. While we describe $B$-conserving and $B$-violating
interactions within the same EFT, 
Refs. \cite{Dover:1981zj,AlvarezEstrada:1982wg}
use phenomenological wavefunctions to represent the deuteron
and pion interactions for two-nucleon $B$-violating corrections.
The strong-interaction pion-nucleon vertex is assumed
to be pseudoscalar instead of pseudovector. 
The latter conforms to chiral symmetry directly, as 
it gives a derivative pion interaction appropriate for a 
(pseudo) Goldstone boson.
The former vertex satisfies chiral symmetry only when it is accompanied
by a seagull vertex, as can be seen by a field redefinition
--- see, for example,
Ref. \cite{Coon:1986kq}. 
The seagull ensures ``pair suppression'' of the so-called $Z$ diagrams.
When the seagull is not included,
pseudoscalar coupling can give anomalously large results,
for example in pion-nucleon scattering and
the closely related pion production in nuclear collisions \cite{Cohen:1995cc}.
An explicit example of this spurious enhancement
involving BSM physics is the neutron
electric dipole moment \cite{Seng:2014pba}. 
Not surprisingly, the result of Ref. \cite{AlvarezEstrada:1982wg}
is very sensitive to the introduction of a  
form factor in the pion-nucleon vertex: a 
reduction factor $\sim 5$ for phenomenological forms.
Moreover, in the absence of a form factor 
the result of Ref. \cite{Dover:1981zj} depends sensitively
on the deuteron wavefunction at small distances: 
a reduction of $\sim 20$ when a hard core is added 
to a Hulth\'en-type wavefunction.

In contrast, our result Eq. \eqref{eqGammdmore} is free of these 
inconsistencies, but our estimate of the magnitude of nuclear corrections relies
on Eq. \eqref{eqImC0}. Since it connects two $S$ waves, this LEC contains an 
enhancement of two powers of $\Lambda_{N\!N}/\kappa$ associated
with the shallowness of the deuteron. Apart from this low-energy enhancement, 
its size estimate is based on NDA. At the one-nucleon level NDA is not
inconsistent with lattice data, see Eq. \eqref{eqGammaest}, but it has not been
tested for $B$ violation at the two-nucleon level.
Eventually $\mathrm{Im}\,C_0$ can be determined by matching 
Eq. \eqref{eqGammdmore} to the deuteron decay rate calculated in lattice QCD.
The same value of the LEC can then be used in the calculation of the decays of 
heavier nuclei.

The low binding energy of the deuteron allowed for a perturbative treatment of 
pion exchange and, as a consequence,
for an analytical calculation in EFT. This calculation is based on
an enhanced scaling of $N\!N$ couplings
--- which absorb the effect of pions with momentum
above the scale $\Lambda_{N\!N}$ ---
with respect to NDA.
The same framework can be used for other light nuclei, where
the binding energy per particle is relatively small --- for a review,
see Ref. \cite{Hammer:2019poc}.
For heavier nuclei, with a larger binding energy per nucleon,
pion exchange might not be amenable to perturbation theory.
Even though the calculation with non-perturbative pions cannot be done 
analytically, we can still apply power counting arguments 
to get an estimate of the relative importance of various contributions.
In this case the typical momentum is $Q \sim \Lambda_{N\!N}$ and 
there is no low-binding enhancement in contact interactions.
With NDA, we find that the decay is still dominated by free-nucleon decay. 
Non-zero corrections from  $\mathrm{ Im}\,C_0$ and from one-pion exchange 
are now expected to appear at the same order,
that is, relative  ${\cal O}(\Lambda_{N\!N}^2/\Lambda_\chi^2)\sim 10^{-2}$. 
Therefore, to a good accuracy, experiments attempting to 
detect $\Delta B=1$ nuclear decay rates can be interpreted as $\Delta B=1$ 
nucleon-decay measurements. 

\section*{Acknowledgments}

We thank Wouter Dekens for useful discussions. 
This research was supported in part
by the Dutch Organization for Scientific Research (NWO)
under program 156 (FO, RGET)
and
by the U.S. Department of Energy, Office of Science, Office of Nuclear Physics,
under award number DE-FG02-04ER41338 (UvK). 
JdV is supported by the 
RHIC Physics Fellow Program of the RIKEN BNL Research Center.

\end{document}